# ChatGPT "contamination": estimating the prevalence of LLMs in the scholarly literature

Andrew Gray

UCL Library Services, University College London,
Gower Street, London WC1E 6BT, United Kingdom

andrew.gray@ucl.ac.uk – ORCID: 0000-0002-2910-3033

## Abstract

The use of ChatGPT and similar Large Language Model (LLM) tools in scholarly communication and academic publishing has been widely discussed since they became easily accessible to a general audience in late 2022. This study uses keywords known to be disproportionately present in LLM-generated text to provide an overall estimate for the prevalence of LLM-assisted writing in the scholarly literature. For the publishing year 2023, it is found that several of those keywords show a distinctive and disproportionate increase in their prevalence, individually and in combination. It is estimated that at least 60,000 papers (slightly over 1% of all articles) were LLM-assisted, though this number could be extended and refined by analysis of other characteristics of the papers or by identification of further indicative keywords.

## LLMs in the scholarly literature

Large language model (LLM) tools provide a way to generate large amounts of reasonably high-quality text automatically, in response to human prompts. While these tools have been available for some years, the release of ChatGPT 3.5 in late 2022, offering an easy interface and wide publicity for the tool, made them available and visible to the world at large. Since then, there has been extensive discussion about their use in scholarly communications. Early over-enthusiasm that they might be able to reliably replace most forms of serious writing, including attempts to list them as authors on papers,[1] has been replaced by a more nuanced understanding of their capabilities and their – substantial – limitations.

As of late 2023, surveys found that 30% of researchers had used tools to help write manuscripts, and many publishers had begun to offer guidance on their use.[2] For example, Wiley permit the use of tools to develop content but only if the authors take full responsibility for the statements made, and where this usage is disclosed transparently in the paper – but "Tools that are used to improve spelling, grammar, and general editing are not included in the scope of these guidelines".[3]

---

[1] See for example Stokel-Walker, Chris (2023). "ChatGPT listed as author on research papers: many scientists disapprove". *Nature*. https://www.nature.com/articles/d41586-023-00107-z
[2] Prillaman, McKenzie (2024). "Is ChatGPT making scientists hyper-productive? The highs and lows of using AI". *Nature*. https://www.nature.com/articles/d41586-024-00592-w
[3] Wiley Author Services (2023). "Best Practice Guidelines on Research Integrity and Publishing Ethics" https://authorservices.wiley.com/ethics-guidelines/index.html

The use of LLMs to generate papers was widely predicted following the release, particularly in "paper mills" producing large amounts of sub-standard articles.[4] However, it is challenging to assess the effect at scale. Some of the distinctive elements of ChatGPT (such as its known propensity to "hallucinate" references[5]) are difficult to identify other than with case-by-case analysis, and – we would hope – are likely to be caught by editors in any case. A number of cases have appeared where fully LLM-generated papers have been identified through the use of obviously non-human phrases such as "As an AI language model…" – but these are relatively rare, a few dozen cases in comparison to the several million papers published each year. It has also been acknowledged that removing particular giveaway phrases can prevent straightforward detection.[6]

Individual "AI detection" tools can work to some degree; for example, analysis of papers published in arXiv has identified a strong uptick in those identified by public "AI detection" tools as being potentially AI-generated; the figure for potentially AI-generated text in computer science preprints rose from around a 3% baseline in 2019 through to a peak of over 7% in late 2023, alongside a growth in total preprint numbers. Physics and mathematics did not show a comparable growth, either in the "potentially AI" proportion or in absolute numbers.[7] However, these tools rely on a detailed analysis of each individual paper.

Recent work (Liang et al 2024)[8] has identified strong evidence that ChatGPT and similar are being used by researchers to generate peer reviews for conference papers in the field of artificial intelligence. Interestingly, they found no indication of a similar pattern in peer reviews for the *Nature* portfolio of journals. Whilst their work was complex and sophisticated, it highlighted some startlingly simple patterns – some common adjectives were found to be used ten to thirty times more in reviews dated 2023, after the public release of ChatGPT 3.5, than in previous years.

## Identification of LLM-associated terms

The words identified by Liang et al suggested a straightforward and simple method of identifying potentially LLM-assisted papers: to look for the terms which were known to be disproportionately associated with LLM-generated text. While this would certainly not indicate with confidence whether any particular use of a term came from such a source, a significant uptick in the use of terms could be a signal of increased use of LLM-generated text in the literature as a whole. It is a much simpler process, not requiring any intensive processing or full-text analysis; we can simply look for the prevalence of these markers throughout the literature.

---

[4] Kendall, Graham; Teixeira da Silva, Jaime A. (2024). "Risks of abuse of large language models, like ChatGPT, in scientific publishing: Authorship, predatory publishing, and paper mills". *Learned Publishing*, 37(1):55-62 doi:10.1002/leap.1578
[5] Buchanan, Joy; Hill, Stephen; Shapoval, Olga (2024). "ChatGPT Hallucinates Non-existent Citations: Evidence from Economics". *The American Economist*, 69(1):80-87 doi:10.1177/05694345231218454
[6] Conroy, Gemma (2023). "Scientific sleuths spot dishonest ChatGPT use in papers". *Nature*. https://www.nature.com/articles/d41586-023-02477-w
[7] Akram, Arslan (2024). "Quantitative Analysis of AI-Generated Texts in Academic Research: A Study of AI Presence in Arxiv Submissions using AI Detection Tool". *arXiv*. https://arxiv.org/abs/2403.13812
[8] Liang, Weixin, et al (2024). "Monitoring AI-Modified Content at Scale: A Case Study on the Impact of ChatGPT on AI Conference Peer Reviews". *arXiv*. https://arxiv.org/abs/2403.07183

While ChatGPT was released to the public in late 2022, the natural delays in the publishing process meant that we would expect only papers with a publication date in 2023 to start showing the effects.

The tables from Liang et al were analysed and the twelve most "unusually frequent" adjectives and adverbs were extracted to test. A set of twelve neutral words were also created which would be expected to appear in a large number of papers, in order to provide a control and an approximate indicator of the share of papers with full text. These words were chosen to be subject-neutral and not have any particular positive or negative tone – by comparison, it is very striking that the vast majority of the adjectives and adverbs originally identified were strongly positive, with only a small handful of terms having a neutral or negative sense.

The Dimensions database was selected for analysis as it was freely available (allowing for reproducibility) and had a high degree of availability for full-text content; around 75% of its indexed articles and reviews are indexed with full text.[9] An early version of the analysis looked at data in Google Scholar, but this was discarded as the source data is less controlled – it was not possible to reliably filter to only published papers – and the numbers of matching papers were not given consistently.

| Adjectives | Adverbs | Controls |
|---|---|---|
| commendable | meticulously | consider |
| innovative | reportedly | conclusion |
| meticulous | lucidly | furthermore |
| intricate | innovatively | relative |
| notable | aptly | technical |
| versatile | methodically | blue |
| noteworthy | excellently | red |
| invaluable | compellingly | yellow |
| pivotal | impressively | before |
| potent | undoubtedly | after |
| fresh | scholarly | earlier |
| ingenious | strategically | later |

Data was then retrieved from Dimensions on the number of documents matching each keyword in the full-text search. Results were filtered to "article" document types only to ensure consistency; no other filters were applied. Data were gathered between 18th and 22nd March 2024.

A count for all "article" documents, with a blank search, was used as the baseline, and the results were calculated as the percentage of documents per year in which the keyword appeared. This baseline rose steadily from around 3.4 million in 2015 to over 5.3 million in 2023. Data for 2024 was collected but not analysed due to incompleteness.

The percentage of documents matching each term ranged from 0.02% ("lucidly", ~1000 articles/year) to over 50% ("after", ~2.8m articles/year); the control words appeared much more frequently than the adjectives, which in turn were generally more common than the adverbs.

---

[9] Herzog, Christian; Hook, Daniel; Konkiel, Stacy (2020). "Dimensions: Bringing down barriers between scientometricians and data". *Quantitative Science Studies* (2020) 1 (1): 387–395. doi:[10.1162/qss_a_00020](10.1162/qss_a_00020)

Given this high variance, the numbers were normalised to show the relative year-on-year change – eg a share of 2.00% one year then 2.10% the following year would become a +5% change – allowing the terms to be directly compared to one another.

## Single word results

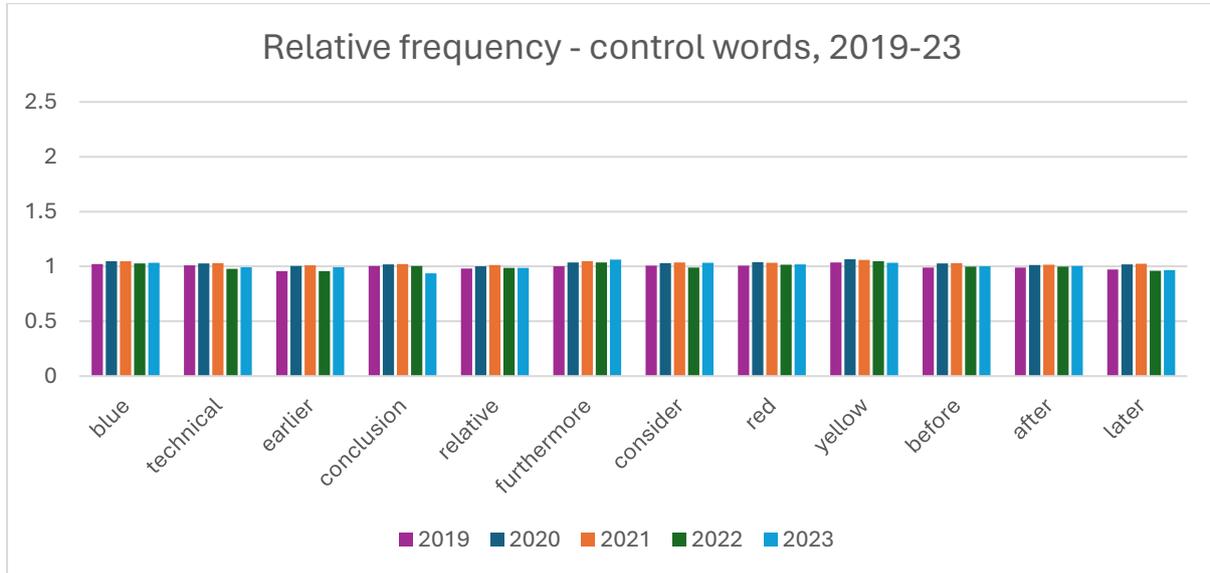

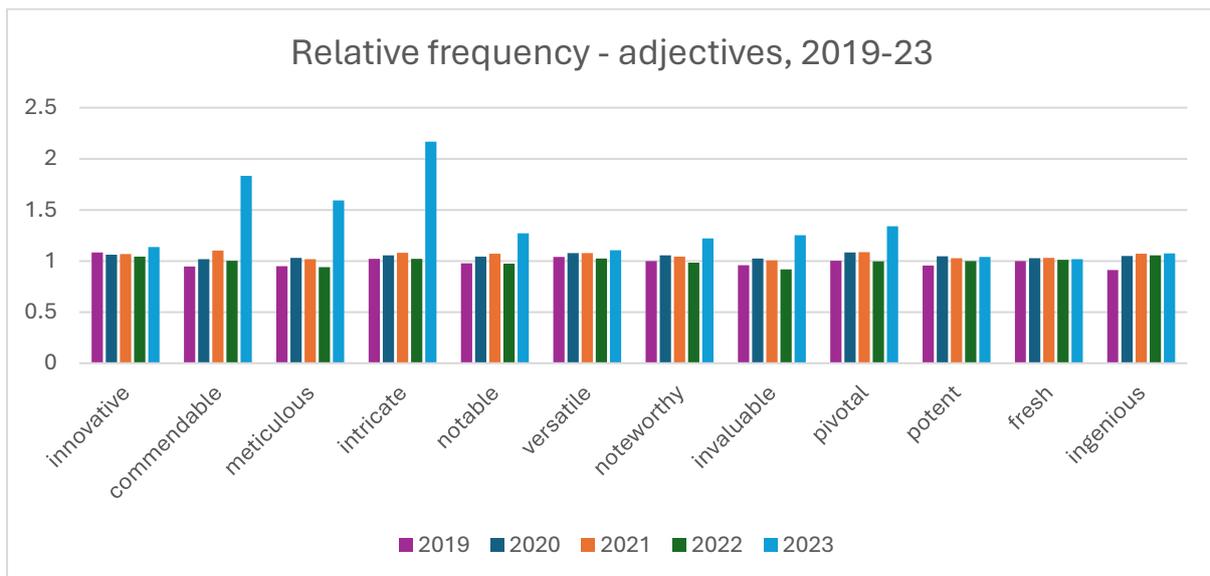

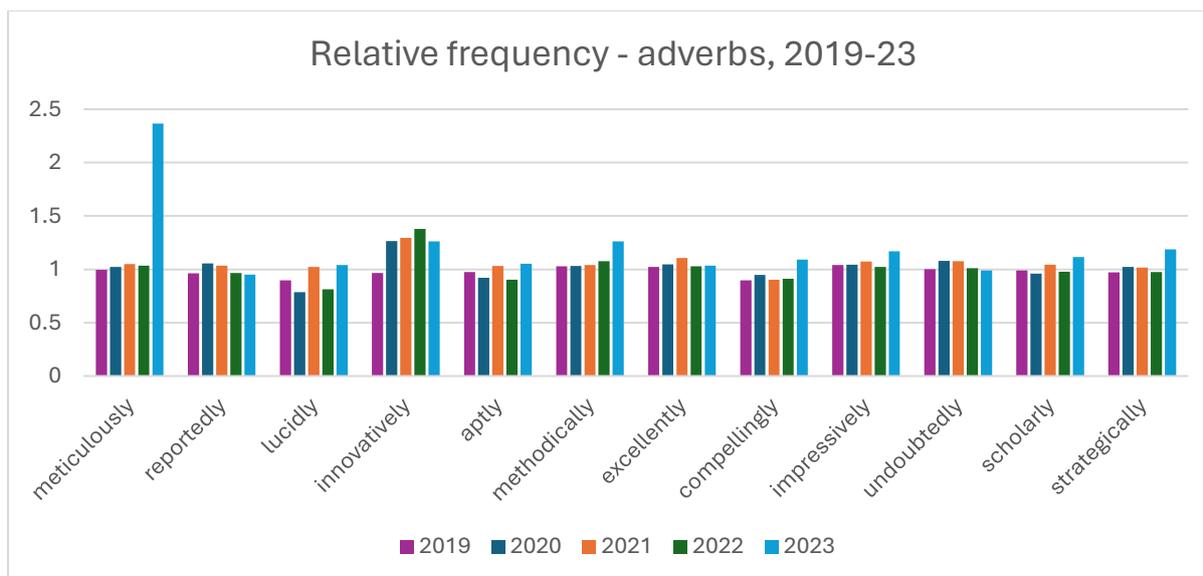

These graphs show the relative frequency change year-on-year for the 36 selected words. The data is only shown for 2019-23 to avoid clutter. Full data for all terms analysed is made available via the UCL Research Data Repository.[10]

As might have been expected, there was very little year-to-year variation among the control terms. Some showed indications of a steady slow growth over time (eg the three colours - "blue" rose from 11.7% in 2015 to 14.5% in 2023; "red" from 16.2% to 18.9%; "yellow" from 5.5% to 7.4%); others were stable or showed a slight decline ("later" fell from 18.8% to 16.6%; "earlier" from 14.7% to 12.0%). These may indicate gradual shifts in preferred wording over time, as different styles of writing become more or less common. The majority of terms did not change by more than ±5% year-on-year, and 2023 was not a particularly unusual year for any of them compared to the preceding years.

These changes may also indicate underlying changes in the number of papers where full-text is available and searchable; it was not possible to baseline against only full-text papers. The low rate of change here is consistent with a relatively consistent level of full-text indexing in each year.

Among the adjectives, things were more complex. Some showed a steady growth 2015-22 ("innovative" and "versatile" averaged about a 5% per year rise), others a slow decline ("meticulous" and "invaluable" averaged about a 3.5% per year decline). The year-on-year changes were a little more volatile than the control terms, which is to be expected given the smaller numbers involved; however, no change was more than 11.3% and most were under ±10%.

The changes in 2023, the "post-LLM year", were very striking. In that year, the twelve adjectives averaged a 33.7% change between 2022 and 2023, with "intricate" rising 117%, "commendable" 83%, and "meticulous" 59%. Eight of the twelve terms had a rise of 10% or higher – noticeably higher than the control words.

The adverbs had greater variability. Some showed a dramatic reduction over the 2015-22 period ("lucidly" averaging a 15% reduction, "compellingly" 9%) while others were stable or increasing ("innovatively" averaging a 25% increase). The yearly variations were often very dramatic – "innovatively" rose almost 60% in one year – perhaps assisted by the relatively small number of papers matching these searches.

---

[10] Gray, Andrew (2024). "LLM related keywords in Dimensions - search counts". University College London. Dataset. doi:10.5522/04/25471957

In 2023, "meticulously" had a 137% increase, and both "methodically" and "innovatively" increased by 26%. The other nine adverbs had a smaller increase or a slight decrease, well within the normal range of variation. One in particular to note is "compellingly", which showed a steady 5-10% per year decline until 2023, when it changed course and began increasing again.

Overall, the adverbs appear less likely to show a 2023 change than the adjectives, and are much less common in general. This is likely due to the different contexts of peer review versus published papers; a review will encourage subjective commentary on the quality of work, but the norms of published papers tend more towards objectivity.

Some words, however, do show a very dramatic year-on-year change in 2023, and have a disproportionate rate of appearance in that year – their relative frequency increases by around 50%. This first group of words are "intricate", "meticulously", "commendable", and "meticulous". A second group of words ("notable", "pivotal", "invaluable", "noteworthy", "methodically", "strategically") show a smaller but still unusual rise. A third pair of words, "innovative" and "versatile", both show clear signs of rising steadily over time, but with a slightly higher than expected increase in 2023. The numbers involved are significant – "intricate", for example, rises from being found in around 50,000 papers per year to well over 100,000.

It appears very likely that these increases are, to some degree, indicative of the widespread use of LLM tools. These words have been independently identified as distinctive markers of the tool, and there is no other likely explanation for a sudden dramatic shift in linguistic style from year to year. No comparable one-year change was noted for most of the tested words in the previous years, with only relatively rare words (eg "lucidly", found in c. 1000 articles/year) showing such dramatic fluctuations.

## Combined terms

The effect is even more striking when combining terms. Articles with one or more of the first group of four "strong" indicators appear 87.4% times more often in 2023; the maximum change in previous years was 6.6%. The second group of "medium strength" indicators show an 18.8% increase against a previous maximum change of 5.4%; the third, "weaker", group show an 11.7% increase against a previous maximum of 7%.

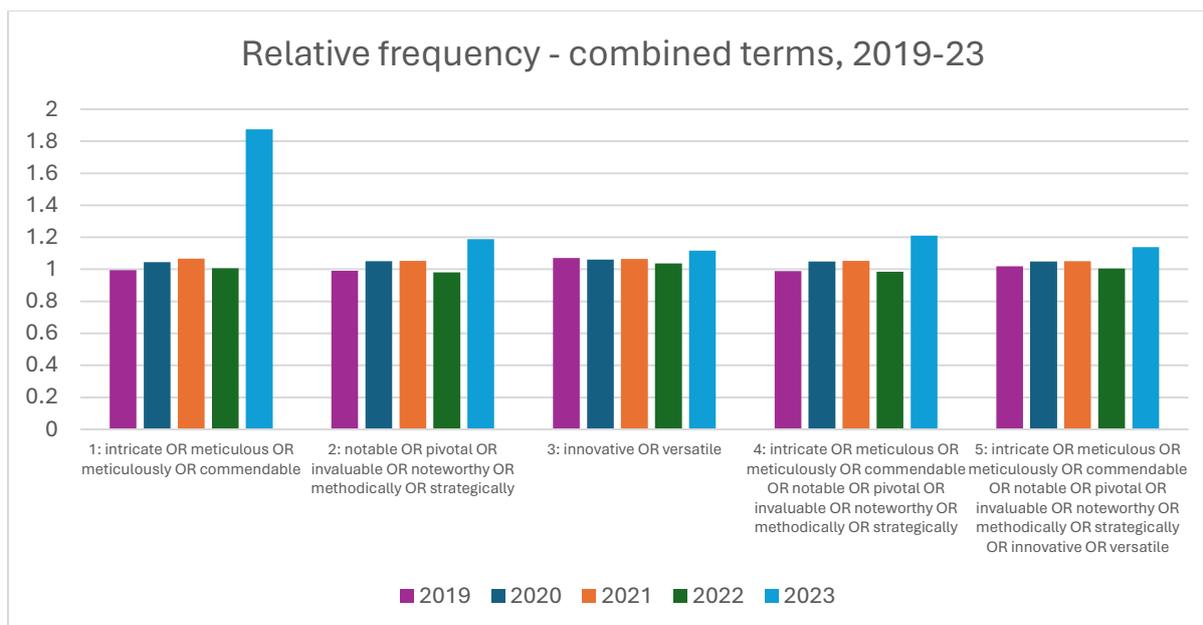

While some of these changes seem proportionally quite low, they are across a very large number of articles – the last combined group of twelve terms represents over one million articles per year, one fifth of all research articles.

|   | Term | 2022 papers | 2023 papers | Increase |
|---|---|---|---|---|
| 1 | **Strong:**<br>intricate OR meticulous OR meticulously OR commendable | 86988 | 159655 | 83.5% |
| 2 | **Medium:**<br>notable OR pivotal OR invaluable OR noteworthy OR methodically OR strategically | 574753 | 668535 | 16.3% |
| 3 | **Weak:**<br>innovative OR versatile | 490213 | 535996 | 9.3% |
| 4 | **Strong & medium:**<br>intricate OR meticulous OR meticulously OR commendable OR notable OR pivotal OR invaluable OR noteworthy OR methodically OR strategically | 634831 | 752334 | 18.5% |
| 5 | **Strong, medium, & weak:**<br>intricate OR meticulous OR meticulously OR commendable OR notable OR pivotal OR invaluable OR noteworthy OR methodically OR strategically OR innovative OR versatile | 1000870 | 1116686 | 11.7% |

It can also be quickly noted that the rate of use of these terms appears to differ by subject area. Across all articles published in 2023, 23.7% are classed as biomedical and clinical sciences, and 14.9% engineering. Among the group 5 articles in the table above, those with one or more LLM keyword marker, however, the figures are 21.9% biomedical/clinical and 22.3% engineering.

## Use of multiple terms

It is reasonable to assume that LLM-generated text which shows a particular fondness for certain words might use them several times, not merely once. Dimensions does not allow us to search for repeated uses of a word in a paper, but it does allow us to look for papers using more than one of our indicator terms. On close examination, the results for certain pairs are very dramatic - for example, the rate of articles with both "intricate" and "meticulous" increases seven-fold; "intricate" and "notable" increases four-fold.

Looking at the frequency of papers that match any combination of two or more terms, we see similar results.

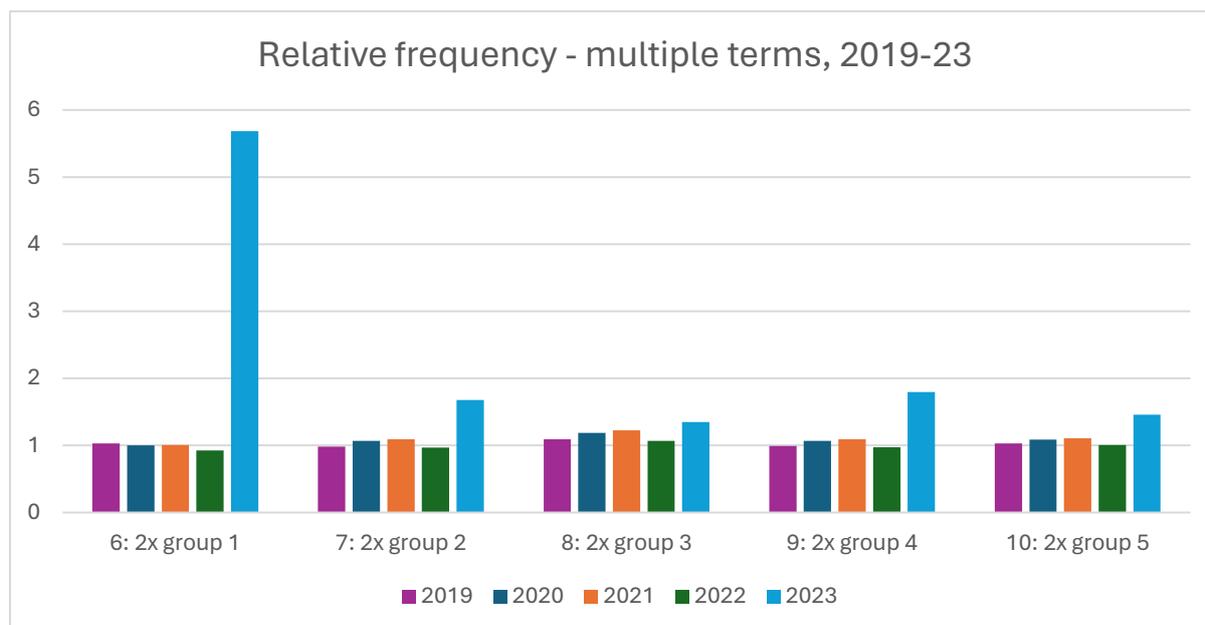

|   | Term | 2022 papers | 2023 papers | Increase |
|---|------|-------------|-------------|----------|
| 6 | **2x group 1 strong terms:** intricate, meticulous, meticulously, commendable | 3045 | 16950 | 468.4% |
| 7 | **2x group 2 medium terms:** notable, pivotal, invaluable, noteworthy, methodically, strategically | 70375 | 115560 | 67.7% |
| 8 | **Both group 3 weak terms:** innovative, versatile | 24807 | 32819 | 35.1% |
| 9 | **2x group 1/2 strong or medium terms** intricate, meticulous, meticulously, commendable, notable, pivotal, invaluable, noteworthy, methodically, strategically | 93002 | 163746 | 79.8% |
| 10 | **2x group 1/2/3 strong, medium, weak terms:** intricate, meticulous, meticulously, commendable, notable, pivotal, invaluable, noteworthy, methodically, strategically, innovative, versatile | 207512 | 296073 | 45.7% |

There is a greater degree of year-to-year fluctuation, particularly among the group 8 (two weak term) results – these had grown by up to 23% in previous years. However, in general, there is still a strong indication that the 2023 results are distinctively and unambiguously higher than the results for earlier years.

## Estimating the overall prevalence

The combined data gives us a first indicator of how many articles, overall, might include LLM derived text. In 2014-22, before LLMs became prevalent, the articles in group 4 (strong + medium terms) averaged an increased frequency of 1.1% per year, and the articles in group 5 (all terms) averaged an increase of 2.1%. The biggest year-on-year change was around 5% for either. As such, a reasonable estimate might be that *in the absence of any external factors*, the number of papers in these groups might have been expected to grow by around 5%. That would give an estimate for group 4 of 666,573 articles, and for group 5 of 1,050,914 papers. The real figures exceeded these by 85,761 and 65,772 respectively – 1.63% and 1.25% of the total number of articles published that year.

If we look at papers using two or more of those terms, groups 9 (two strong/medium terms) or 10 (two strong/medium/weak terms) show a maximum year-on-year increase of around 10-11% in 2014-22, and then rises of 79.8% and 45.7% respectively. Allowing a generously estimated rise of 11% for each group, we would see a projected total of 103,232 papers for group 9, and 230,338 for group 10. The actual figures exceeded these by 60,514 and 65,735 respectively (1.15% and 1.25% of the total).

At the moment, we appear able to say that there is an "excess" of papers in 2023 which use indicative terms that correlate with being, at least, "polished" by a LLM tool. These "excess papers", even allowing a conservative estimate for normal year-on-year growth, represent between 1.15% and 1.63% of the articles recorded by Dimensions in 2023, *over and above* the papers that might be expected to use one of those terms – a total excess between 60,000 and 85,000 papers.

Three additional points should be noted.

Firstly, these words are not the only markers that could be used to identify potentially ChatGPT derived text – for example, "groundbreaking" is presumably unlikely to be used in most peer reviews, but is likely to occur in the literature reviews of papers. It had a 52% increase in 2023, higher than most of the other terms tested. "Outwith", a term normally only used in Scottish English, turns out to be unexpectedly favoured by ChatGPT as well – it almost tripled in 2023, rising 185%. It is very likely that there are other words not tested here which are similarly indicative of "ChatGPT style" and more likely to be found in articles – which could potentially drive the counts further up. Logically, it is likely that there are also *negative* correlations – words which ChatGPT is unlikely to use but a human would be more likely to select. It is likely that some negative-weighted terms may fall into this category. A cursory investigation suggests, for example, that while "inconclusive" and "ineffective" show no change, "incompetent" has shown a slight but noticeable drop in 2023-24.

Secondly, very few papers in this group disclose any use of LLMs. Out of over a million articles in group 5, the broadest set of keywords, only 767 (>0.1%) also matched a text search suggesting disclosure (chatGPT or GPT or openAI or LLM or "large language model" or "artificial intelligence") – and a quick skim of those articles suggests that many are false positives, articles *about* LLM tools rather than those disclosing assistance.

Thirdly, the situation is accelerating. Data for 2024 is obviously as yet incomplete, and may be skewed in various ways, but the share in groups 4/5 (a single keyword) and 9/10 (multiple keywords) shows significant increases for 2024 over and above 2023. While the frequency of these terms was increasing slowly in 2015-22, the effect is out of all proportion to what was seen before. It is hard to interpret this as anything other than a rapidly increased use of LLM text generation and copyediting.

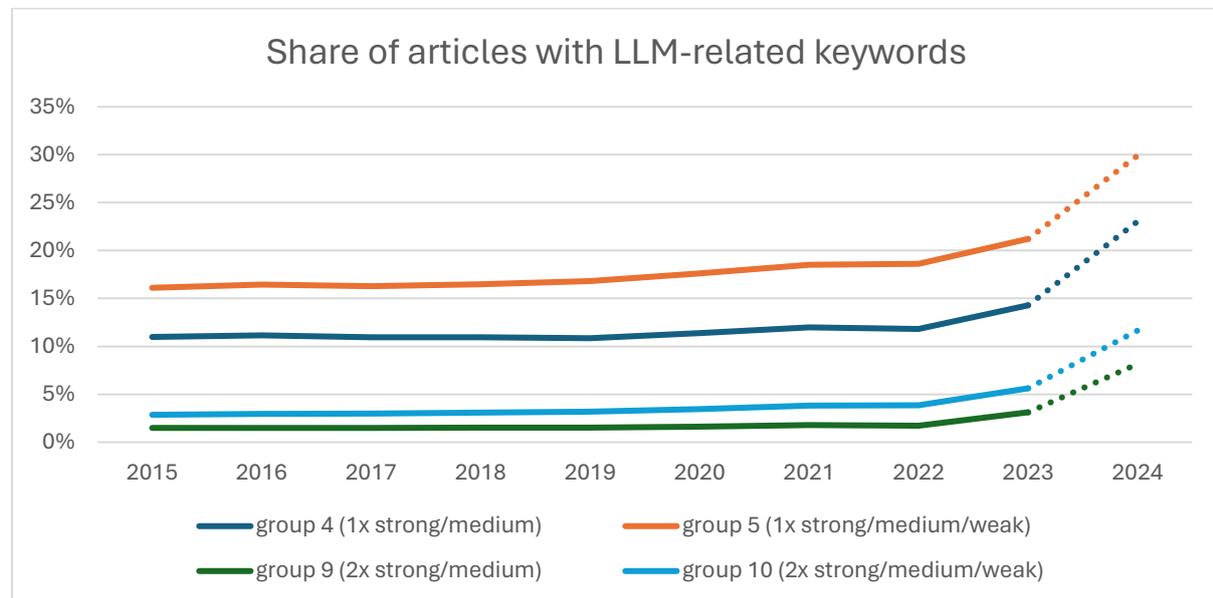

## Is this a problem?

We have seen that it is likely a small but significant amount of papers published in 2023 – perhaps upwards of 60,000 articles – contained a disproportionate occurrence of some unusual words that are known to be correlated with LLM generated text. This does not of course indicate that any *particular* paper was constructed or assisted by an LLM, nor does the absence of these terms necessarily indicate that an LLM was not involved. But it is a remarkably visible effect, and it is difficult to find a conclusion that does not implicate LLM use.

There are two key implications.

The first is the question of whether these tools are merely being used for purely stylistic purposes. It is difficult to make this determination without a more detailed case-by-case analysis, but it seems *plausible* that at least some of the rise does indicate the use of LLMs for more than simple stylistic polishing. The Liang study identified that LLM marker words were often correlated with other factors indicating a less involved author – for example, submission shortly before a deadline, lower confidence in the conclusions, and fewer citations to other works. While none of these can be easily tested with a simple database search, it seems reasonable to assume that the correlations found for peer reviews also hold up for papers – which would imply a substantial proportion of these papers may have more significant LLM involvement than was openly disclosed.

This will need further work to identify the scale of the issue, and more aggressive engagement by publishers and reviewers to push back on the ethical aspects of undisclosed and unsupervised LLM use that go beyond simple copyediting.

The second is the implications for LLMs themselves. The scholarly literature has been a major input source for many LLMs. However, the more LLM-generated text is used as training material for future LLM generations, the higher the risk of "model collapse" – artificially generated text outweighing real text, producing increasingly low-quality results.[11] The more that the scholarly literature contains undisclosed LLM-assisted text, the worse that future LLM-assisted text may become, in a vicious cycle. Indeed, if LLM-assisted text starts to represent a significant fraction of the literature, it may start to influence human linguistic choices as that becomes the "normal style", further complicating matters.

At best, this may just be stylistic – we could end up with the scholarly literature of the 2030s sounding strangely positive and upbeat, with literature reviews endlessly praising the "notable", "intricate" and "meticulous" works of their colleagues. At worst, we could see the quality of the models degenerating in other ways beyond simple stylistic choices, alongside a growing reliance on them – implying that a significant amount of research papers might have significant content issues that authors and publishers are not able to properly identify.

## Future assessment

It seems likely, at the time of writing, that the same analysis performed in a year will show a dramatic and ongoing rise for the assessed keywords through 2024 and into 2025.

This study has only assessed the prevalence of certain keywords, but other characteristics of the papers may well be correlated with markers of LLM assistance. For example, we have quickly tested that it is likely to be correlated with subject area, and this would certainly reward deeper analysis. It may plausibly be correlated with other features such as the number of papers or level of collaboration on a paper – it is easy to hypothesise that, for example, if it only requires one author in a collaboration to push back against using LLM-generated text, then papers with large cross-institution research groups would be less likely to contain such markers. Fully LLM-generated text might be more likely with a single author. There may also be a different pattern of LLM text usage among reviews, or in conference proceedings or monographs, which would be worth investigating further – for example, different publishing contexts might be likely to use different distinctive terms.

This study has only looked at whether the keywords feature at all in a paper – it seems plausible that in-text frequency would also be a marker, with LLM-generated text using the distinctive keyword with different frequencies to human-generated text. Lastly, the known tendency of LLMs to generate nonsensical references would – hopefully – not survive into most published papers, but it would be a useful test to perform on corpuses of preprints or submitted papers.

## Conclusion

We have seen that the use of distinctive terms can be used, at a large scale, to estimate the prevalence of LLM-assisted papers. As a first estimate, this is on the order of 60,000 articles, or slightly over 1% of the scholarly articles published in 2023. It is likely that this estimate could be improved by further investigation of distinctive terms, or by other distinctive characteristics of the papers.

---

[11] Shumailov, Ilia, et al. (2023). "The Curse of Recursion: Training on Generated Data Makes Models Forget". *arXiv* https://arxiv.org/abs/2305.17493

The rate is expected to increase significantly through 2024. Explicit disclosure of the assistance of LLM tools is very rare by comparison, though it should be acknowledged most publishers do not require this for mere copyediting assistance. However, it is difficult to imagine that they are being used *solely* for this purpose. Authors who are using LLM-generated text must be pressured to disclose this – or to think twice about whether doing so is appropriate in the first place – as a matter of basic research integrity. From the other side, publishers may need to be more aggressive to engage with authors to identify signs of undisclosed LLM use, and push back on this or require disclosure as they feel is appropriate.

## Acknowledgments

With thanks to colleagues in UCL Libraries for discussions of the issues around LLMs, and to Bev Ager and Kirsty Wallis for their feedback on early drafts of this paper.